# Atmospheric Turbulence-Immune Free Space Optical Communication System based on Discrete-Time Analog Transmission

Hongyu Huang, Zhenming Yu, *Member, IEEE*, Yi Lei, Wei Zhang, Yongli Zhao, Shanguo Huang, and Kun Xu

*Abstract*—To effectively mitigate the influence of atmospheric turbulence, a novel discrete-time analog transmission free-space optical (DTAT-FSO) communication scheme is proposed. It directly maps information sources to discrete-time analog symbols via joint source-channel coding and modulation. Differently from traditional digital free space optical (TD-FSO) schemes, the proposed DTAT-FSO approach can automatically adapt to the variation of the channel state, with no need to adjust the specific modulation and coding scheme. The performance of the DTAT-FSO system was evaluated in both intensity modulation/direct detection (IM/DD) and coherent FSO systems for high-resolution image transmission. The results show that the DTAT-FSO reliably transmits images at low received optical powers (ROPs) and automatically enhances quality at high ROPs, while the TD-FSO experiences cliff and leveling effects when the channel state varies. With respect to the TD-FSO scheme, the DTAT-FSO scheme improved receiver sensitivity by 2.5 dB in the IM/DD FSO system and 0.8 dB in the coherent FSO system, and it achieved superior image fidelity under the same ROP. The automatic adaptation feature and improved performance of the DTAT-FSO suggest its potential for terrestrial, airborne, and satellite optical networks, addressing challenges posed by atmospheric turbulence.

## I. INTRODUCTION

FREE-space optical (FSO) communication has the advantages of high bandwidth, easy deployment, high security, and desirable cost-effectiveness, being an effective solution for integrated terrestrial, airborne, and satellite networks [1], [2]. However, the transmission performance of FSO is limited by impairments such as geometric loss and pointing error and influences of atmospheric attenuation and turbulence. In particular, the atmospheric turbulence effect will cause an adverse effect known as scintillation or fading, leading to performance degradation especially in long-distance transmissions of several kilometers [3], [4].

Currently, traditional digital free-space optical (TD-FSO) communication systems are often used for point-to-point transmission. In these schemes, information sources (such as text, images, and videos) are first converted into bits via source coding and encoded via channel coding and then mapped to symbols via a modulator for transmission [5]. However, TD-FSO systems suffer from a "cliff effect" in which the performance will drop like a cliff when the channel signal-to-noise ratio (SNR) deteriorates below a threshold [6]. In other words, the performance of digital communication systems critically depends on channel states [7]. However, in FSO communication systems, the SNR varies over time due to the optical intensity scintillation induced by atmospheric turbulence. This will significantly impact the performance of the TD-FSO scheme. To improve system performance, many adaptive coding and modulation schemes have been proposed by adjusting the modulation and coding scheme according to the estimated channel state [8], [9]. As the coherence time of the FSO channel is typically 0.1 to 10 ms, millisecond-level adjustment speed for adaptive coding and modulation schemes is required [10]. This is challenging for a practical system. In addition, the adaptive coding and modulation scheme requires an additional channel information feedback link, which increases system complexity.

In this paper, we propose a free-space optical communication scheme based on discrete-time analog transmission (DTAT-FSO), with the aim of enhancing FSO transmission performance under atmospheric turbulence. With no bit-based coding or modulation, the proposed DTAT-FSO scheme can effectively avoid the "cliff" and "leveling" effects. In particular, a joint source-channel coding and modulation (JSCCM) is proposed to directly map the information sources to channel symbols (while in the traditional system, this function is completed using source coding, channel coding, and modulation separately). These channel symbols, discrete in the time domain with continuous amplitude, characterize

Manuscript received × × ××; revised × × ××. This work was financially supported by the National Key R&D Program of China (No. 2023YFB2905900); the National Natural Science Foundation of China (No. 61821001, 61901045); and the Fund of State Key Laboratory of Information Photonics and Optical Communications, BUPT (No. IPOC2021ZT18). Supported by BUPT Excellent Ph.D. Students Foundation (CX2023136). (Corresponding author: Zhenming Yu.)

Hongyu Huang, Zhenming Yu, Wei Zhang, Yongli Zhao, Shanguo Huang, and Kun Xu are with State Key Laboratory of Information Photonics and Optical Communications, Beijing University of Posts and Telecommunications, Beijing, China (e-mails: hongyuhuang@bupt.edu.cn; yuzhenming@bupt.edu.cn; zw_bupt@bupt.edu.cn; yonglizhao@bupt.edu.cn; shghuang@bupt.edu.cn; xukun@bupt.edu.cn).

Yi Lei is with the School of Computer Science and Information Engineering, Hefei University of Technology, Hefei, China (e-mail: leiyi@hfut.edu.cn).

the system as discrete-time analog transmission[11], [12].

The main contributions of this study are summarized as follows:
- A discrete-time analog transmission scheme is proposed for the first time for FSO communication.
- A transformer-based JSCCM network is designed for image transmission; it constrains the peak-to-average power ratio (PAPR) of the output symbols to ensure an efficient transmission in physical optical links.
- The performance of the DTAT-FSO scheme is evaluated in intensity modulation/direct detection (IM/DD) and coherent FSO systems both, showing high transmission quality under atmospheric turbulence effects.

Due to the characteristics of discrete-time analog symbols, the DTAT-FSO exhibits a graceful and gradual performance degradation similar to analog communication when the received optical power (ROP) decreases. In the presence of atmospheric turbulence, the DTAT-FSO demonstrates automatic adaptation to channel noise variation, whereas the TD-FSO requires adjustment in coding and modulation schemes. Moreover, the DTAT-FSO shows superior image transmission performance. When compared to the TD-FSO scheme, the DTAT-FSO enhances receiver sensitivity by more than 2.5 dB in the IM/DD FSO system and by 0.8 dB in the coherent FSO system. Under identical ROP conditions, the DTAT-FSO scheme achieves higher image transmission fidelity, notably enhancing fidelity in textures and details.

## II. CHALLENGES AND MOTIVATIONS

### A. Challenges of TD-FSO

Fig. 1 illustrates the structure of the TD-FSO and DTAT-FSO systems. In the TD-FSO structure, source coding, channel coding, and modulation are designed separately. Source encoding compresses information, while channel coding adds redundancy to correct bit errors during channel transmission. Modulation converts the encoded bits into symbols for transmission. In the TD-FSO systems, forward error correction (FEC) codes such as Reed–Solomon codes, polar codes, and low-density parity check codes (LDPC) are commonly considered for channel coding [8], [13], [14]. These FEC coding schemes exhibit a threshold effect, wherein the performance of system will rapidly decline when the bit error rate after channel transmission (before FEC decoder) exceeds a certain threshold (i.e., the number of errors exceeds the error-correcting capability of the FEC codes). Bit error seriously affects the quality of the received information. Therefore, the TD-FSO system based on a separate design faces two challenges:
- First, when system SNR varies rapidly under the effect of atmospheric turbulence, the TD-FSO system is prone to experiencing mismatches between the system design and the channel state. Two inherent issues in digital transmission systems are the cliff and leveling effects. The cliff effect occurs when the system's BER beyond the FEC threshold as the channel condition varies, leading to a rapid increase bit error probability. This failure in the source decoding process results in a substantial decrease in reconstruction quality. Similarly, the leveling effect is observed when the system cannot benefit from enhanced channel conditions either. Specifically, when source and channel coding rates are fixed, no matter how good the channel is, the reconstruction quality remains the same as long as the channel capacity is above the target rate [6], [15].
- Second, based on Shannon's theorem, the separation of source and channel coding can achieve optimality as the block length approaches infinity. Nevertheless, to ensure system latency, the block length is limited in practice, especially in applications with low-latency demands. Furthermore, the optimality of separation design breaks down with non-ergodic source or channel distribution [16], [17]. Hence, communication systems based on separate design usually exhibit a performance gap compared to the theoretical limits.

### B. Motivation for DTAT-FSO

The structure of the proposed DTAT-FSO is depicted in Fig. 1. A notable feature is the extraction of the desired features of information sources and mapping them into channel symbols directly using a JSCCM network. This process eliminates the need for bit-based coding and modulation. The performance of the DTAT-FSO system shares similarities with analog communication system, where the quality of source reconstruction closely depends on the channel state. As a result, the DTAT-FSO system can effectively avoid the cliff and leveling effects and adapt to the channel state variation automatically. Considering the time-varying nature of the channel state under atmospheric turbulence, the DTAT-FSO system based on integrated design is a promising approach to enhancing the system's performance.

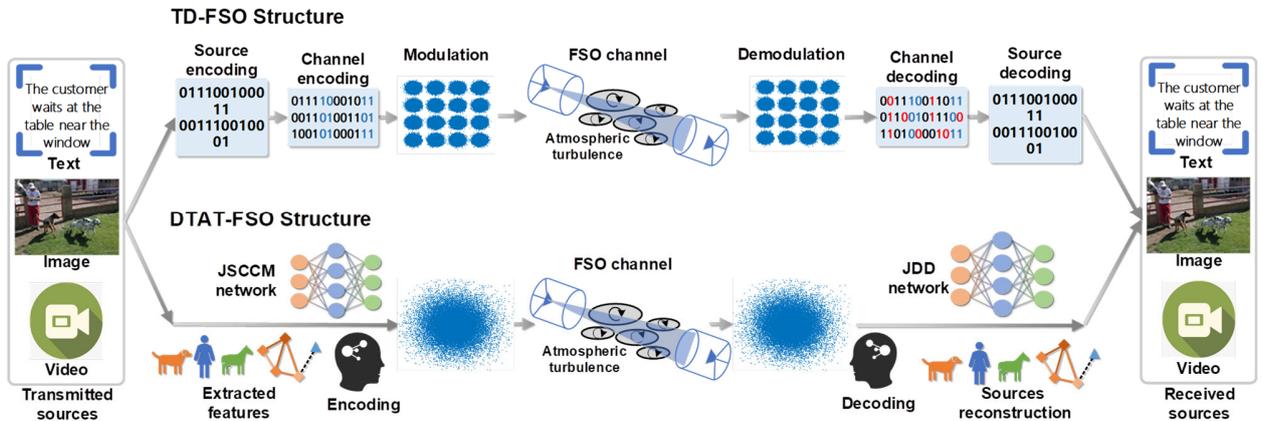

**Fig. 1.** Structure of TD-FSO and DTAT-FSO systems.

## III. THE PROPOSED DTAT-FSO SYSTEM

### A. JSCCM Network

In the DTAT-FSO, information sources are mapped into channel symbols by the JSCCM network. Then, the symbols are transmitted through the FSO channel. After transmission, the symbols are demodulated and decoded by the joint demodulation decoding (JDD) network. The JSCCM network is the key part of the DTAT-FSO system. The following description focuses on designing the JSCCM for high-resolution image transmission in our DTAT-FSO system. Inspired by advanced artificial intelligence technology, deep convolution neural networks (CNN) and transformer structures are widely utilized for image encoding. Limited by model capacity, the performance of a CNN-based network degrades rapidly with an increase in image dimensions. For encoding high-resolution images, transformer-based networks show greater potential [18]–[20]. To achieve high-performance transmission of high-resolution images, the designed JSCCM network is based on the Swin Transformer [21], as shown in Fig. 2. The RGB images $x \in R^{H \times W \times 3}$ are split into $l_1 = (H/2) \times (W/2)$ non-overlapping patches. The patches are input into the network in order from top left to bottom right. After patch embedding, the Swin Transformer blocks are utilized for encoding. After undergoing four stages of encoding and a fully connected layer, the images are encoded into discrete-time analog symbols. The constellation of the DTAT-FSO system is shown in Fig. 2; it is irregular and has a near-Gaussian distribution of amplitude. After transmission through the FSO channel, the images are recovered by the JDD network, which acts as the inverse process of the JSCCM.

During training, images from the DIV2K dataset were used [22]. These images were cropped randomly into 256 × 256 patches. The training batch size was set to 16. We employed an Adam optimizer with a learning rate of $1\times10^{-4}$. To enhance the noise robustness of the DTAT-FSO system, the JSCCM and JDD network models were trained end-to-end under an additive white Gaussian noise channel, with a uniform distribution for SNRs ranging from 1 to 14 dB.

Having a reasonable signal PAPR is important for transmission over the physical FSO link, as a large PAPR will lead to distortions caused by nonlinear devices like AD/DA converters and modulators [23]. To ensure good performance in image encoding and simultaneously control the PAPR of the output discrete-time analog symbols, we design a loss function $L$ for the DTAT-FSO model training,

$$L = (1 - M_s) + \lambda P , \quad (1)$$

where $M_s$ is a multi-scale perceptual metric (MS-SSIM) that closely approximates human visual perception, assessing the performance of the received images [24]. $P$ indicates the PAPR of the output symbols. The term $\lambda P$ limits the PAPR of the output discrete-time analog symbols, where $\lambda$ is a tunable parameter. After optimization, $\lambda$ was set to $5 \times 10^{-4}$ in this study.

### B. FSO Channel Model

Inhomogeneities in temperature and pressure within atmosphere, attributed to solar heating and wind dynamics, induce variations in the air refractive index along the transmission path. These variations, termed atmospheric turbulence, result in random fluctuations in optical power. Various stochastic channel models have been proposed to characterize the distribution of turbulence-induced fading in FSO systems. These include the log-normal distribution, negative exponential distribution, double Weibull distribution, and gamma–gamma distribution models. Among them, the gamma–gamma model demonstrates good performance across weak to strong fluctuation regimes [2], [3]. Therefore, in this study, the gamma–gamma model was selected to model atmospheric turbulence effects [4].

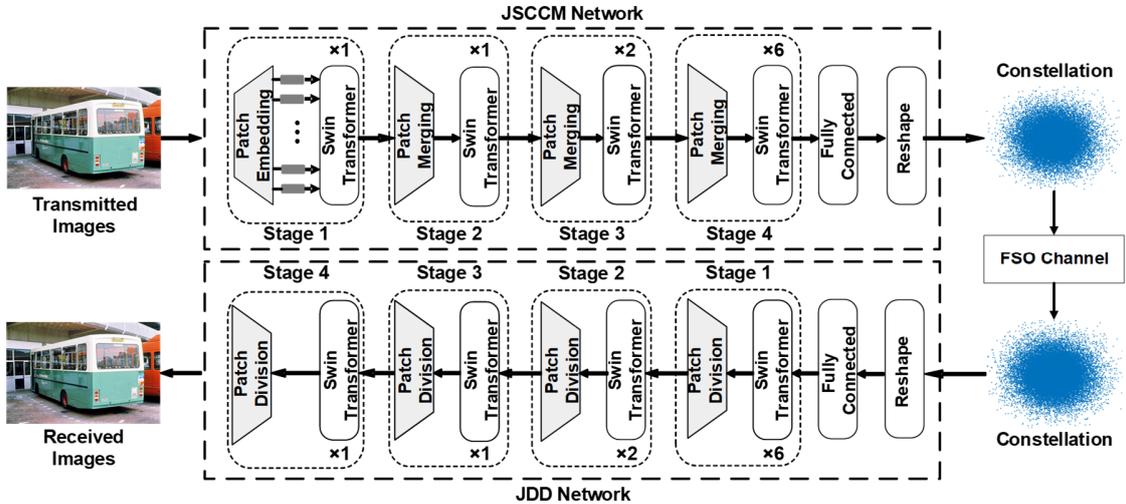

**Fig. 2.** Principle of the DTAT-FSO system.

In the gamma–gamma model, the probability density function of the received intensity $I$ is

$$P(I) = \frac{2(\alpha\beta)^{(\alpha+\beta)/2}}{\Gamma(\alpha)\Gamma(\beta)} I^{(\alpha+\beta)/2-1} K_{\alpha-\beta}(2\sqrt{\alpha\beta I}) \,, \quad (2)$$

where $\Gamma(.)$ is the gamma function and $K_{\alpha-\beta}(.)$ is the modified Bessel function of the second kind; $\alpha$ and $\beta$ characterize the intensity of the scintillation index and are defined as

$$\alpha = [0.49\sigma_R^2 / (1+1.11\sigma_R^{12/5})^{7/6} - 1]^{-1}, \quad (3)$$

$$\beta = [0.51\sigma_R^2 / (1+0.69\sigma_R^{12/5})^{5/6} - 1]^{-1}. \quad (4)$$

Further, $\sigma_R^2$ can be calculated as follows:

$$\sigma_R^2 = 1.23 C_n^2 k^{7/6} Z^{11/6}, \quad (5)$$

where $C_n^2$ is the refractive index structure parameter and $k$ is the wavenumber.

## IV. EVALUATION IN IM/DD SYSTEM

Both IM/DD and coherent structures are commonly used in FSO transmission systems. IM/DD offers advantages in simplicity and cost-effectiveness, while coherent architecture enables higher transmission rates and improves receiver sensitivity. The performance of the proposed DTAT-FSO system was evaluated in IM/DD and coherent FSO systems both. We initially implemented an IM/DD FSO system to evaluate the transmission performance of the DTAT-FSO. Fig. 3 illustrates the system setup, where high-resolution images are transmitted to assess the DTAT-FSO scheme's performance. The system's baud rate and sampling rates are 10 Gbaud and 20 GSa/s, respectively. The channel symbols generated from coding and modulation are processed by the digital signal processing at the transmitter (Tx-DSP). The Tx-DSP includes the square root raised cosine (SRRC) filtering with a 0.1 roll-off factor and resample. Then, the electrical signals are modulated using the Mach-Zehnder modulator (MZM). The optical amplifier amplifies the transmitting optical power. After transmission through the FSO channel, the optical signals are converted to electrical signals by the photodiode (PD). These electrical signals are processed by the digital signal processing at the receiver (Rx-DSP), which includes resample, matched filtering, and a 61-tap forward feedback equalization (FFE).

A TD-FSO system with advanced source and channel coding was employed for comparison. Specifically, the JPEG2000 (JP2K) and 3/4 code rate LDPC coding were used as source coding and channel coding, respectively [25], [26]. The modulation format included OOK and PAM4.

With a fixed system baud rate, the image transmission rate of the system relies on the number of symbols obtained after encoding and modulation for each image. To enable a fair comparison of transmission performance at a similar image transmission rate, the TD-FSO and DTAT-FSO schemes produced a comparable number of symbols after encoding and modulation for each image. High-resolution images from the Kodak dataset and ADE20k with $768 \times 512 \times 3$ pixels were transmitted [27], [28]. For a single image with $768 \times 512 \times 3$ pixels, the number of bits and symbols obtained after encoding and modulation are described in Table I. The image transmission rate $R$ can be calculated by

$$R = R_s / N_{SPI}, \quad (6)$$

where $R_s$ is the symbol rate and $N_{SPI}$ is the number of symbols per image after coding and modulation.

TABLE I
NUMBER OF BITS AND SYMBOLS OBTAINED AFTER ENCODING AND MODULATION

| Coding and modulation scheme | Number of bits after JPEG2000 | Number of bits after LDPC | Number of symbols after modulation |
|---|---|---|---|
| JPEG2000+3/4-rate LDPC+OOK (TD-FSO) | 112345 | 149796 | 149796 |
| JPEG2000+3/4-rate LDPC+PAM4 (TD-FSO) | 224801 | 299736 | 149868 |
| | **Number of symbols after JSCCM** | | |
| JSCCM (DTAT-FSO) | 147456 | | |

According to the data in Table I, the image transmission rates of the TD-FSO and DTAT-FSO were approximately $6.68 \times 10^4$ and $6.78 \times 10^4$ images/s, respectively. The net bit rates of the OOK and PAM4 transmissions were 7.5 Gb/s and 15 Gb/s, respectively. By adjusting the compression ratio of the source coding, the image transmission rates of the TD-FSO and DTAT-FSO were controlled to be similar. The JP2K encoding of the OOK transmission required a higher source compression rate.

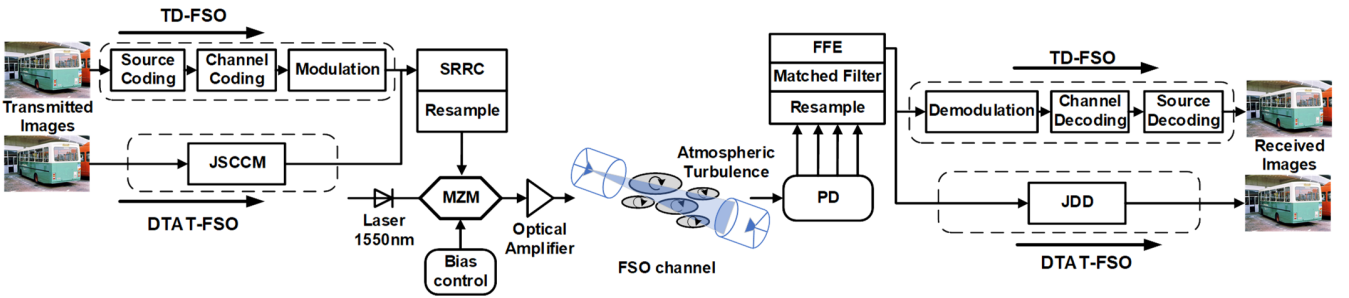

**Fig. 3.** Setup of the IM/DD FSO system.

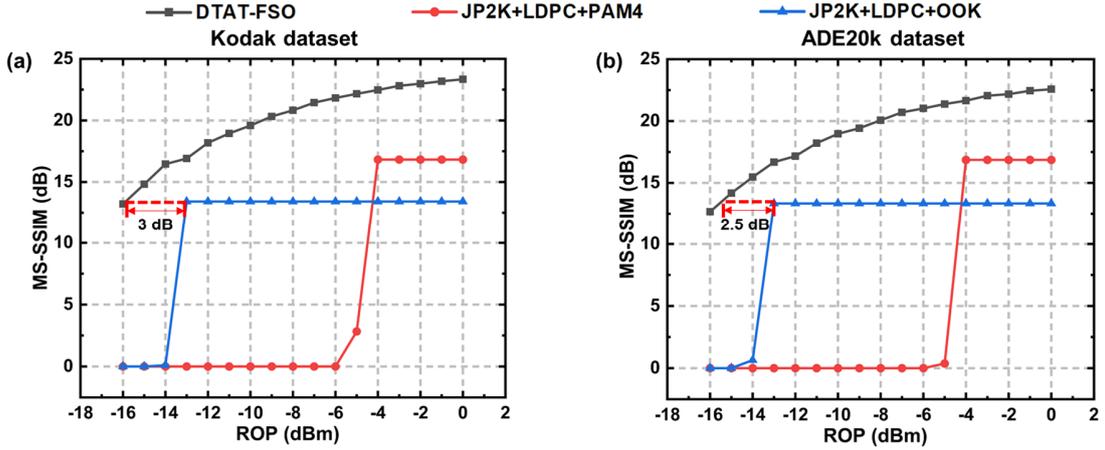

**Fig. 4.** Relationships between ROP and transmission performance of (a) Kodak dataset and (b) ADE20k dataset in the IM/DD FSO system.

The receiver sensitivities of the various transmission schemes were investigated in an optical back-to-back (OB2B) transmission link. The ROP was controlled using an optical attenuator. The relationships between the ROP and transmission performance of the Kodak dataset and ADE20k dataset are depicted in Fig. 4. For more intuitive observation and comparison, the MS-SSIM is converted into dB form. The formula is MS-SSIM (dB) = −10log(1 − MS-SSIM). In the TD-FSO system, there is a trade-off between the quality of source coding and the robustness of channel impairments. At the same baud rate, using a high-order modulation format (PAM4 is considered in this paper) enables the transmission of more bits, supporting higher quality source transmission. However, a high-order modulation format implies poorer robustness to channel impairments, compared to low-order modulation (OOK). When the ROP is below −5 dBm, the PAM4 transmission exhibits a "cliff" effect. For the OOK transmission, the receiver sensitivity reaches about −13 dBm. However, with increasing ROP, the OOK transmission may not efficiently utilize the channel, and the quality of image transmission remains the same. In contrast, channel noise will impact the quality of the received discrete-time continuous amplitude symbols in the DTAT-FSO transmission, directly influencing the performance of the received image. Therefore, the DTAT-FSO can automatically adapt to variations in ROP. As the ROP decreases, the DTAT-FSO provides a graceful and gradual performance degradation similar to analog communication. Meanwhile, the DTAT-FSO improves the transmission quality of the image significantly, achieving over 2.5 dB receiver sensitivity improvement in the low ROP region.

After the OB2B transmission, the performance of the DTAT-FSO was investigated in a 5 km FSO channel with atmospheric turbulence effects. The parameters of the FSO channel are shown in Table II. The refractive index structure parameter $C_n^2 (m^{-2/3})$ was set to $1\times10^{-15}$ (moderate turbulence). The coherence time of the FSO channel was 1 ms. In the FSO system, where the transmission rate was very high (10 Gbaud), the channel fading coefficient remained constant over millions of symbols. Therefore, the channel fading was quasi-static [3].

TABLE II
PARAMETERS OF THE FSO CHANNEL

| Parameter | Value |
| --- | --- |
| Carrier wavelength | 1550 nm |
| Attenuation | 0.443 dB/km |
| Beam divergence | 0.25 mrad |
| Transmitter aperture diameter | 5 cm |
| Receiver aperture diameter | 20 cm |
| Scintillation model | gamma–gamma |
| Index refraction structure | 1e-15 |

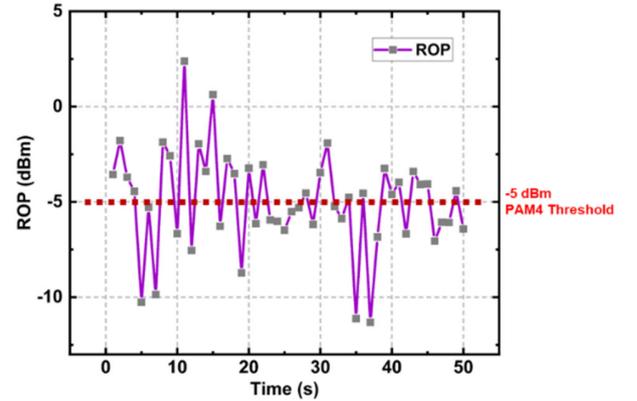

**Fig. 5.** Measured time-varying ROP of the FSO system.

The transmitting optical power was set to 15 dBm. We performed data collection 50 times at intervals of 1 second. The relationship between measured ROP and time is illustrated in Fig. 5. Due to the atmospheric turbulence effect, the ROP was time-varying. When the ROP was below −5 dBm, reliable transmission of PAM4 became challenging, as demonstrated in the OB2B transmission. The information could not be transmitted effectively at 40% of the sampling time in PAM4 transmission.

The transmission results in the FSO channel are shown in

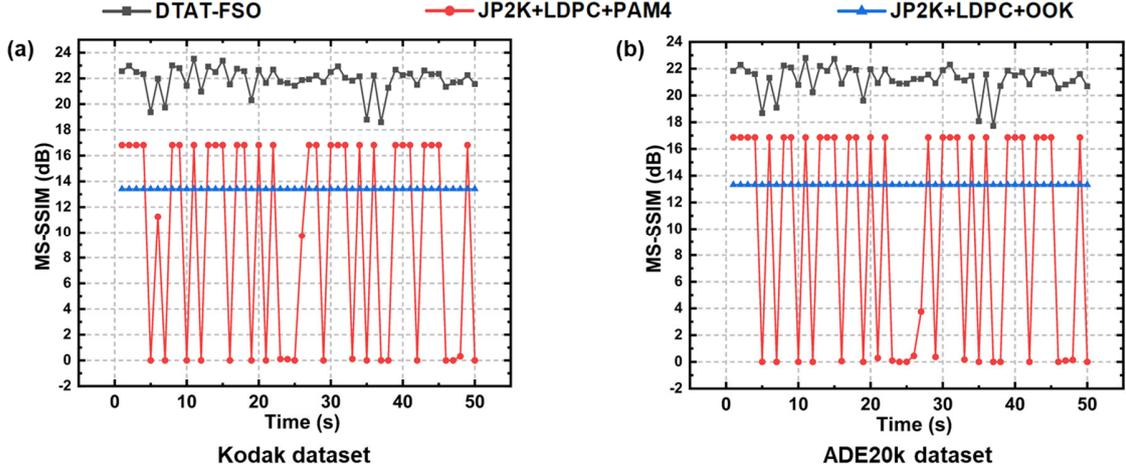

**Fig. 6.** Transmission performance of the IM/DD FSO system for (a) Kodak dataset and (b) ADE20k dataset.

Fig. 6. The PAM4 transmission scheme experienced unreliable communication with fluctuations in ROP. Specifically, when the ROP was low, the receiver did not effectively receive the transmitted information. The OOK transmission achieved reliable transmission throughout all sampling times. However, it did not enhance transmission performance at high ROPs. In comparison, the DTAT-FSO demonstrated automatic adaptation to ROP variations. It achieved reliable transmission at low ROPs and automatically enhanced image transmission quality at high ROPs. Furthermore, the overall performance of the DTAT-FSO transmission surpassed that of TD-FSO. The visual results of the fifth sampling are shown in Fig. 7. At the fifth sampling, the ROP was about −10.5 dBm. The PAM4 scheme did not transmit the image efficiently. Moreover, the DTAT-FSO achieved superior visual quality compared to the OOK-based TD-FSO scheme. Specifically, the DTAT-FSO produced higher fidelity textures and details.

V. EVALUATION IN COHERENT SYSTEM

We also conducted a dual-polarization coherent FSO system to investigate the transmission performance of the DTAT-FSO scheme. Fig. 8 shows the system setup, where high-resolution images were transmitted to assess the DTAT-FSO system's performance. The system baud rate and sampling rate were 25 Gbaud and 50 GSa/s, respectively. The channel symbols generated from coding and modulation are processed by the Tx-DSP, which includes SRRC filtering with a 0.1 roll-off factor and resample. Then, the electrical signals are converted into optical signals by the dual-polarization IQ modulator (DP-IQ modulator). The optical signals are then amplified. After transmission through the FSO channel, the optical signals are converted to electrical signals by the coherent receiver. The electrical signals are processed by the Rx-DSP.

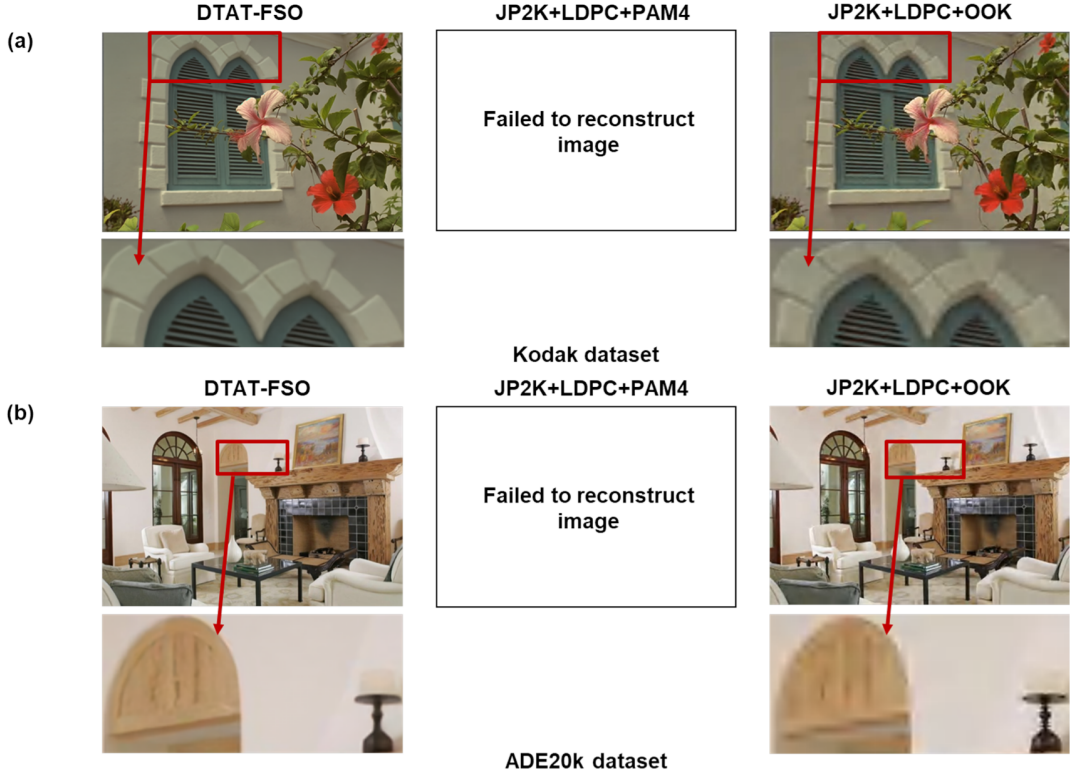

**Fig. 7.** Visual results at the fifth sampling: (a) Kodak dataset and (b) ADE20k dataset.

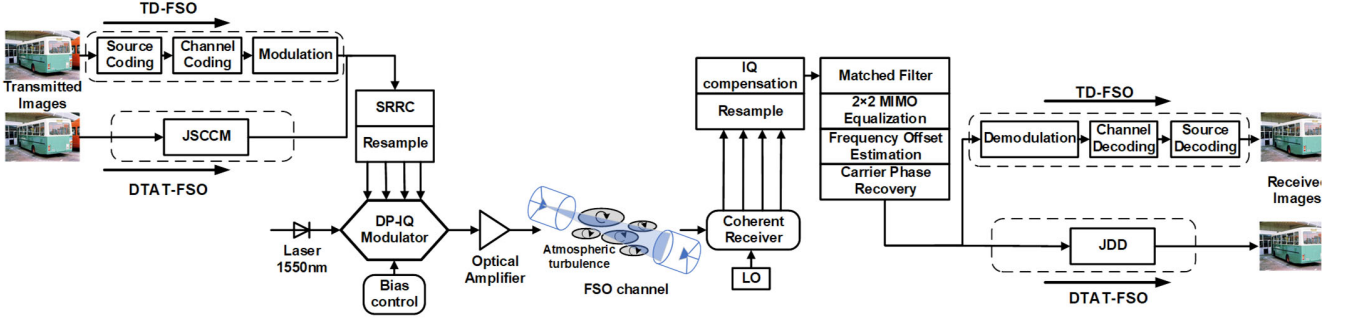

**Fig. 8.** System setup of the coherent FSO system.

A 2 × 2 multiple-input multiple-output (MIMO) equalizer was employed for depolarization and channel equalization. Since the constellation of the DTAT-FSO is irregular, a pilot-aided algorithm was used for carrier phase recovery (CPR) [29], [30]. In our coherent DTAT-FSO system, the pilot overhead was 2%. The coding schemes for the coherent TD-FSO were identical to those used in the IM/DD TD-FSO system. The modulation schemes of the coherent TD-FSO were QPSK and 16QAM. For the DTAT-FSO, the discrete analog symbols outputted by the JSCCM network were transformed from serial to parallel into four channels and converted into optical signals by DP-IQ modulation. The net bit rates of QPSK and 16QAM transmission were 75 Gb/s and 150 Gb/s, respectively. The image transmission rate of the coherent TD-FSO system was $6.67 \times 10^5$ images/s. The image transmission rate of the coherent DTAT-FSO was $6.65 \times 10^5$ images/s, where 2% pilot overhead was considered.

The receiver sensitivities of different transmission schemes were investigated in OB2B coherent transmission. The ROP was regulated by an optical attenuator. The relationships between ROP and transmission performance of both the Kodak dataset and ADE20k dataset in the coherent system are shown in Fig. 9. Similar to the results obtained in the IM/DD FSO system, the DTAT-FSO obtained superior image transmission quality. The DTAT-FSO adapted the channel state variation automatically, thereby avoiding the cliff effect. Moreover, the receiver sensitivity was improved by over 0.8 dB using DTAT-FSO. This improvement was slightly less than that seen with the IM/DD FSO system. This is because of the irregular constellation of DTAT-FSO, which limits the performance of the CPR algorithm.

The performance of the coherent DTAT-FSO system was investigated in a 5 km FSO channel. The parameters of the channel were consistent with those of the IM/DD FSO system mentioned in Section IV-B. The transmitting optical power was set to 10 dBm. We conducted data collection 50 times at intervals of 1 second. The transmission results of the coherent FSO system are shown in Fig. 10. Under the influence of the atmospheric turbulence effect, the 16QAM transmission scheme suffered unreliable communication. The 16QAM transmission frequently failed to effectively transmit the information. The QPSK transmission achieved reliable transmission at all sampling times, but it was unable to improve transmission performance at high ROPs. In comparison, the DTAT-FSO automatically adapted to ROP variations. The DTAT-FSO achieved reliable transmission at low ROPs. At high ROPs, it transmitted the image with high fidelity. Meanwhile, the overall image transmission quality of the DTAT-FSO was better than that of the TD-FSO. The visual results of the 17th sampling are shown in Fig. 11. At the 17th sampling, the ROP was about −9.7 dBm. Both the TD-FSO and the DTAT-FSO schemes efficiently transmitted the image. (As can be seen in Fig. 10, the receiver sensitivities of the 16QAM and QPSK transmissions were −19 dBm and −12 dBm, respectively.) The DTAT-FSO scheme obtained the best visual results among the schemes, with higher fidelity textures and details. In the TD-FSO scheme, the visual results based on 16QAM transmission were better than those based on QPSK transmission.

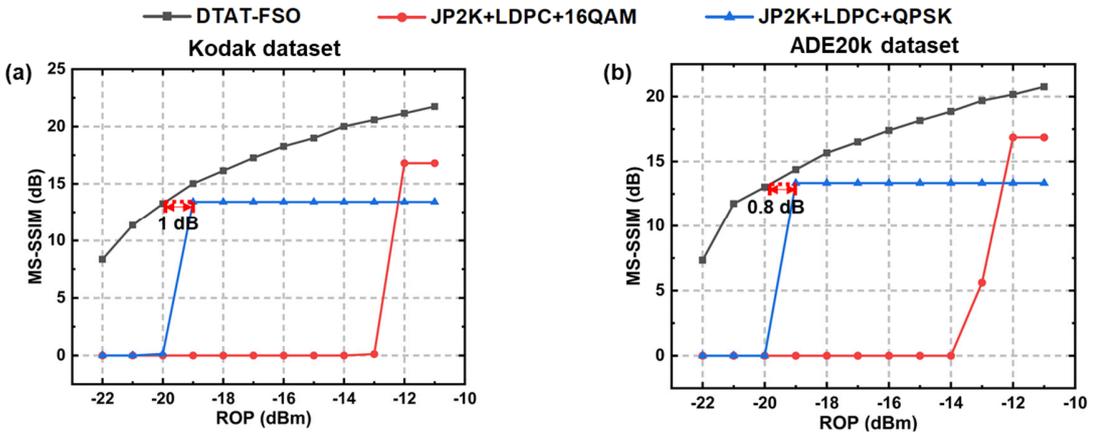

**Fig. 9.** Relationships between ROP and the transmission performance of (a) Kodak dataset and (b) ADE20k dataset in the coherent system.

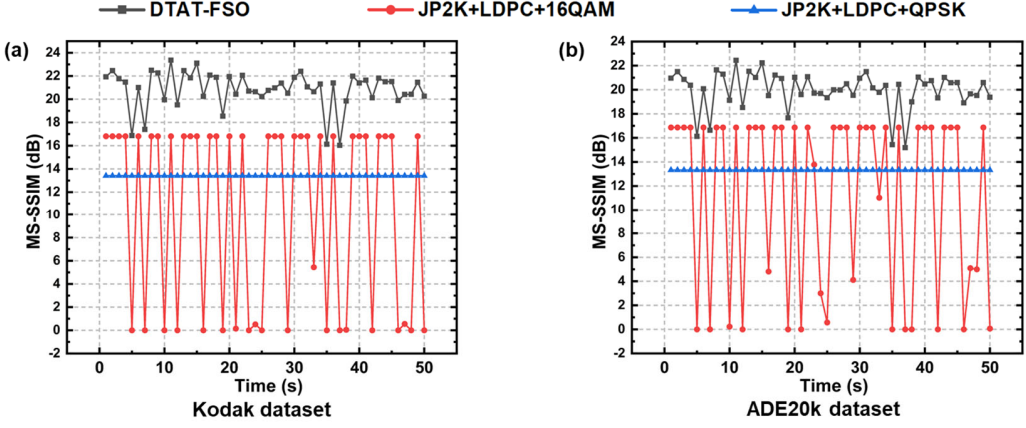

**Fig. 10.** Transmission performance of the coherent FSO system for (a) Kodak dataset and (b) ADE20k dataset.

## VI. CONCLUSION

In this paper, we propose a novel free-space optical communication system based on discrete-time analog transmission. In the DTAT-FSO system, the deep learning-based JSCCM network maps the information source directly to the channel symbols, without traditional source coding, channel coding, and modulation. These channel symbols are discrete in the time domain with continuous amplitude. We established both IM/DD and coherent FSO systems and compared the performance of the DTAT-FSO with traditional digital FSO schemes. High-resolution images were transmitted to evaluate the performance. The DTAT-FSO scheme showed excellent capability in automatically adapting to channel state

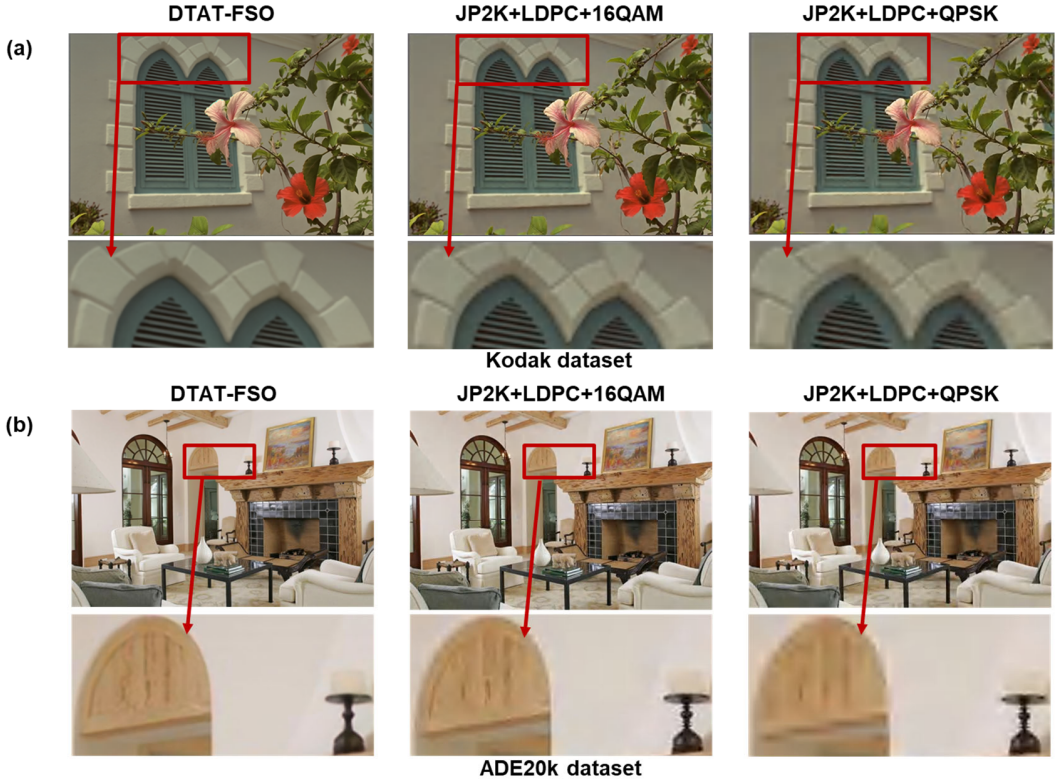

**Fig. 11.** Visual results at the 17th sampling: (a) Kodak dataset and (b) ADE20k dataset.

variations and enhancing information transmission quality.

In scenarios with time-varying FSO channels affected by atmospheric turbulence, designing separated digital encoding and modulation schemes becomes challenging. If the separated scheme is designed for high impairment robustness, the information transmission ability under favorable channel conditions will be limited. Conversely, if the separated scheme is designed for high information transmission ability, it will face communication interruptions during low-power optical signal reception. In contrast, the DTAT-FSO reliably transmits information at low ROPs and enhances transmission quality automatically at high ROPs. Moreover, adoption of the DTAT-FSO scheme results in improvements in the system's receiver sensitivity, with enhancements of over 2.5 dB in the IM/DD FSO system and 0.8 dB in the coherent FSO system. The automatic adaptation feature and improved performance of the DTAT-FSO hold promise for advancing the development of terrestrial, airborne, and satellite networks by offering improved reliability and information transmission rates in FSO systems.

In addition to the image source, the DTAT-FSO scheme can support effective transmission of other information sources through the design of appropriate JSCCM networks; it will then improve the application capabilities of FSO system in sensing, positioning, multimedia, and other areas.